\documentclass[conference]{IEEEtran}
\usepackage{cite}
\usepackage{amsmath,amssymb,amsfonts}
\usepackage{algorithmic}
\usepackage{float}
\usepackage{graphicx}
\usepackage{hyperref}
\usepackage{textcomp}
\usepackage{xcolor}
\usepackage[frozencache]{minted}

\def\BibTeX{{\rm B\kern-.05em{\sc i\kern-.025em b}\kern-.08em
    T\kern-.1667em\lower.7ex\hbox{E}\kern-.125emX}}
\definecolor{mannibg}{HTML}{f0f3f3}

\definecolor{mannibg}{HTML}{f0f3f3}

\begin{document}

\title{Research Project 2: Drone-supported AI-based Generation of 3D Maps of Indoor Radio Environments}

\author{
	\IEEEauthorblockN{Ken Mendes}
	\IEEEauthorblockA{\textit{Department of Computer Science} \\
		\textit{University of Antwerp}\\
		Antwerp, Belgium \\
		ken.mendes@student.uantwerpen.be}
}

\maketitle

\begin{abstract}
A Radio Environment Map (REM) is a powerful tool in enhancing the experience of radio-enabled agents but building such a REM can be a laborious undertaking, especially in three dimensions. This project shows how such a REM of an indoor three-dimensional space  can be generated in an autonomous and scalable way. Building on the results of the preceding Research Project 1, multiple drones are used to map the WiFi signals present in such a space in a real-world environment where the drones are each able to visit 36 waypoints and collectively gather thousands of WiFi beacon data samples. This report also includes an analysis of the collected data and concludes by proposing machine-learning based techniques to predict the signal strength of known access points in locations not visited by the drones.
\end{abstract}

\begin{IEEEkeywords}
drone, autonomous, REM, Radio Environment Map, Indoor, Crazyflie, ESP8266
\end{IEEEkeywords}

\section{Introduction}
This research project is a continuation of \textit{Research Project I: Drone-based Autonomous
Generation of 3D Maps of Indoor Radio Environments} \cite{project1} where we showed how a customized Crazyflie drone can autonomously gather IEEE 802.11b, 802.11g and 802.11n beacon frame data in the 2.4 GHz Industrial, Scientific and Medical (ISM) band and send that data to a base station for storage and further processing. The drone was equipped with a custom integrated WiFi module and an Ultra Wide Band (UWB) positioning system with high accuracy. The drones got their instructions, including waypoints to visit, from software running on a base station which featured the ability to strategically shut down the controlling radio to minimize interference while the drone scans the IEEE 802.11 2.4 GHz band.

Building on that work, this project aims to provide a framework for using additional drones that can be seamlessly integrated into the system, allowing for sequential data collection with multiple drones. This results in a scalable way to efficiently gather beacon frame data within a room while making improvements to the drones' stability when scanning compared to the previous project. As before, we are primarily interested in collecting the signal strength of access points in the vicinity. Once this data is collected, further analysis will be done to show the possibilities this method provides on a larger scale.

The drones gather this data at discrete locations within a three-dimensional space. We conclude this project by proposing machine-learning models that can predict the signal strength of these recorded access points with a reasonable error at positions within that space which were not visited.

\section{Scalability and enhancements}
\subsection{Drones}
\begin{figure}[H]
	\begin{center}
		\includegraphics[width=\columnwidth]{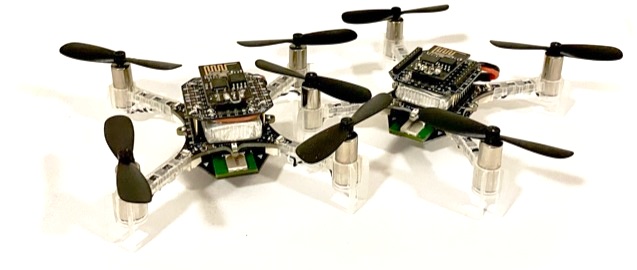}
		\caption{Two customized Crazyflie 2.1 drones}
		\label{fig:two_crazyflie_drones}
	\end{center}
\end{figure}

This project uses two Crazyflie 2.1 \cite{crazyflie-2.1} drones as shown in figure \ref{fig:two_crazyflie_drones} for the data collection. 

The hardware configuration of these drones has not changed, they are each equipped with two expansion decks:

\begin{itemize}
	\item A Loco Positioning Deck that acts as a tag in the UWB based Loco Positioning System (LPS), this allows the drones to calculate their position with decimetre-level accuracy
	\item A custom deck with an ESP8266 WiFi module used to collect IEEE 802.11b, 802.11g and 802.11n signals in the 2.4 GHz ISM band
\end{itemize}

To benefit from upstream fixes and improvements, the custom firmware \cite{esp8266-cf-firmware} of the drones was rebased on the 2021.06 \cite{crazyflie-firmware-2021.06} \textit{crazyflie-firmware} release.

\subsection{Hovering improvements}
During project 1, one of the major measures we took to minimize radio interference was to shut down the controlling radio during the three seconds that measurements were being collected by the WiFi module. Both operate in the 2.4 GHz ISM band and the reduction in interference was significant.

Unfortunately, we also saw some occasional drifting of the drone during those few seconds that it lost contact with the base station. When the drone loses its radio connection, it also loses its ability to get new setpoints (target positions) from the base station. When no new setpoint is received for over 500 ms, the drone will set its attitude angles (pitch, roll and yaw) to 0 in order to keep itself stabilized. While this system does provide some stability, it is not necessarily making the drone hold its position when it has sufficient momentum or for example an unbalanced propeller.

\begin{figure}[h]
	\begin{center}
		\includegraphics[width=\columnwidth]{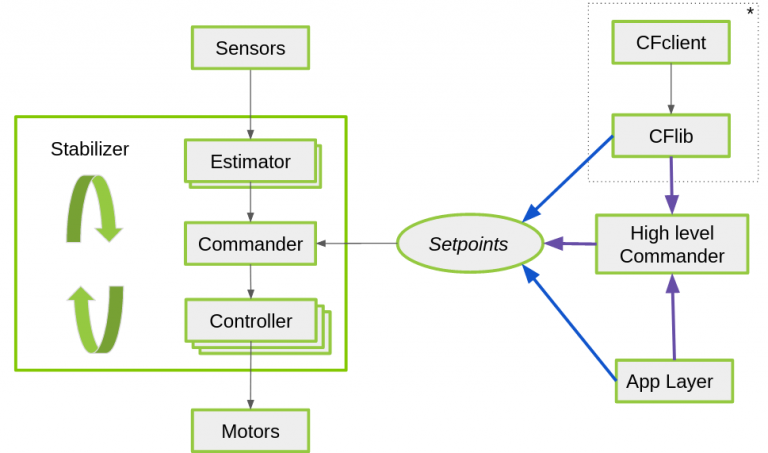}
		\caption{The Crazyflie commander framework}
		\label{fig:commander_framework}
	\end{center}
\end{figure}

Figure \ref{fig:commander_framework} details how the base station's custom Python client can forward setpoints (target locations) to the \textit{Commander} in the drone's firmware by making use of the \textit{CFlib} library.

In order to make the drone hold its position after shutting down the radio connection, an extra FreeRTOS Task \ref{listing:1} was added to the driver of the ESP8266 deck that will feed back the scanning position every 100ms to the drone's commander during such a scan. This task gets resumed at the start of the scanning task and suspended at the end of it so that it doesn't interfere with regular waypoint activities. This feedback process results in the drone not just having stability while scanning, but also actively maintaining its position.

\begin{listing}[h]
\caption{Hovering Task}
\begin{minted}[style=tango,bgcolor=mannibg,fontsize=\footnotesize]{C}
static void hoverWhileScanning(void* arg) {
  setpoint_t hoverSetpoint;
  uint8_t hoverCount = 0;
	
  while(1) {
    // Hover while we are scanning
    getHoverSetpoint(
      &hoverSetpoint, 
      position.x, 
      position.y, 
      position.z
    );

    if (hoverCount == 10) {
      consolePrintf(
        "Hovering at position x: %f, y: %f, z: %f\n",
        (double)hoverSetpoint.position.x, 
        (double)hoverSetpoint.position.y, 
        (double)hoverSetpoint.position.z
      );
      
      hoverCount = 0;
    }
		
    commanderSetSetpoint(&hoverSetpoint, 3);
    hoverCount++;
    vTaskDelay(M2T(100));
  }
}
\end{minted}
\label{listing:1}
\end{listing}

\section{Data collection}
\subsection{Test environment} \label{subsec:test-environment}
The 3D volume for the drones to scan is a rectangular cuboid of 3.74m long (x-axis), 3.20m wide (y-axis) and 2.10m high (z-axis), located in a living room in a big apartment building. This provides a real-world environment where signals of many access points in different configurations are available. 

At each of the 8 corners of the cuboid, an LPS anchor is placed to enable the drone to calculate its position within the volume.

\begin{figure}[h]
	\begin{center}
		\includegraphics[width=\columnwidth]{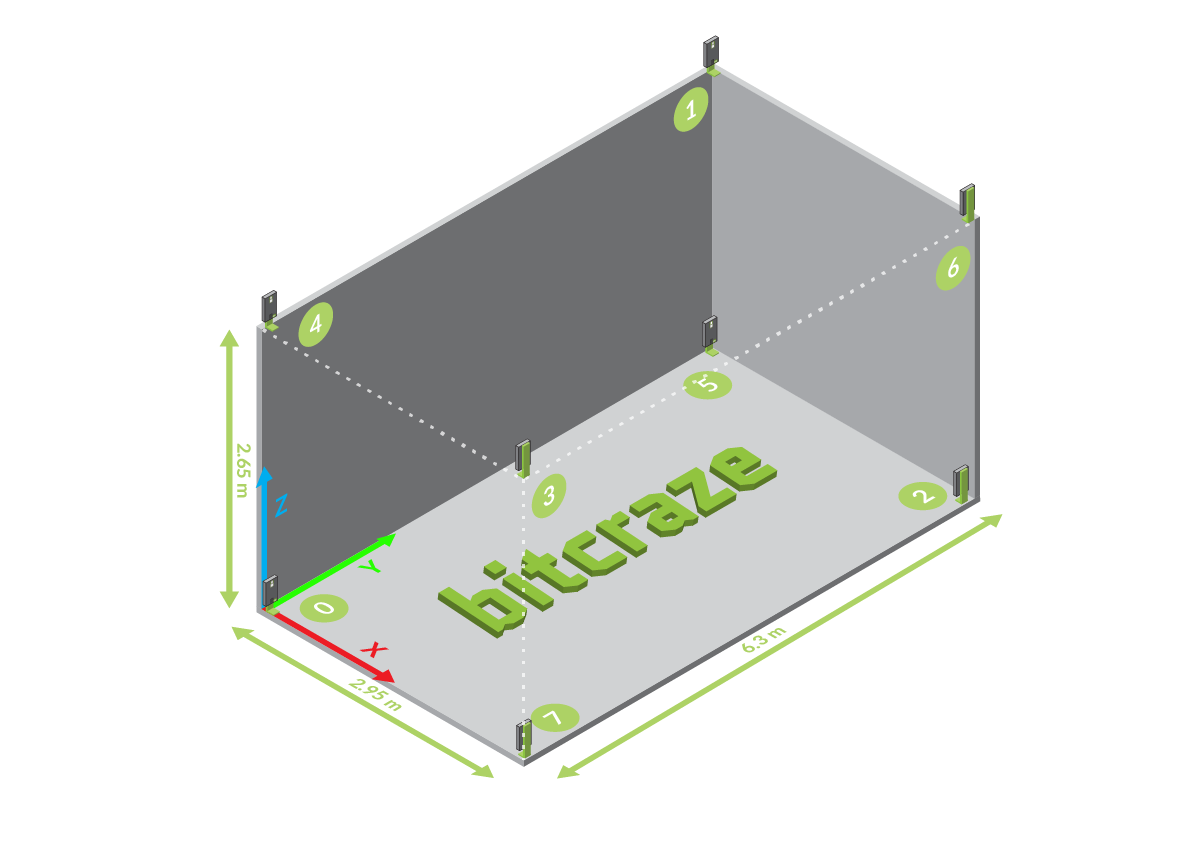}
		\caption{8 Anchors LPS reference setup}
		\label{fig:crazyflie_lps_reference_setup}
	\end{center}
\end{figure}

Anchor placement is done according to the 8 anchor reference setup of Bitcraze as shown in figure \ref{fig:crazyflie_lps_reference_setup}. 

In order to minimize radio interference, the drones are run in sequence, not in parallel. Since we're working with multiple drones, the Loco Positioning System is configured to use one of the Time Difference of Arrival protocols (TDoA2) instead of the Two Way Ranging (TWR) protocol.

Positions of the anchors are documented in table \ref{tab:anchor_positions}.

\begin{table}[h]
	\begin{center}
		\begin{tabular}{| c | c | c | c |}
			\hline
			Anchor ID & x & y & z \\ \hline
			\hline
			$0$ & $0.00$ & $0.00$ & $0.00$ \\ \hline
			$1$ & $0.00$ & $2.30$ & $2.10$ \\ \hline
			$2$ & $3.74$ & $2.31$ & $0.00$ \\ \hline
			$3$ & $3.74$ & $0.00$ & $2.09$ \\ \hline
			$4$ & $0.00$ & $0.00$ & $2.10$ \\ \hline
			$5$ & $0.00$ & $2.33$ & $0.00$ \\ \hline
			$6$ & $3.74$ & $2.30$ & $2.09$ \\ \hline
			$7$ & $3.74$ & $0.00$ & $0.00$ \\ \hline
		\end{tabular}
	\end{center}
	\caption{LPS Anchor positions in meters}
	\label{tab:anchor_positions}
\end{table}

\subsection{Scan locations}
The endurance test we ran during project 1 \cite{project1} showed that a Crazyflie drone in this configuration can operate for a little over 6 minutes reliably when flying stationary and performing a scan every 8 seconds. We can expect its endurance to be lower when the demands increase: visiting different locations and doing more frequent scans.

With this constraint in mind, 72 locations evenly spread over the volume to scan were identified with each drone being responsible for scanning 36 of them. The drones have 4 seconds to fly from one location to the next and require 3 seconds to perform a scan. Scanning 36 locations should therefore take at least $36 \times (4s + 3s) = 252s$ or 4 minutes and 12 seconds. If we add the time required to take off, land and the more intensive itinerary, the drones will come close to their maximum operating time. 

\begin{figure}[h]
	\begin{center}
		\includegraphics[width=\columnwidth]{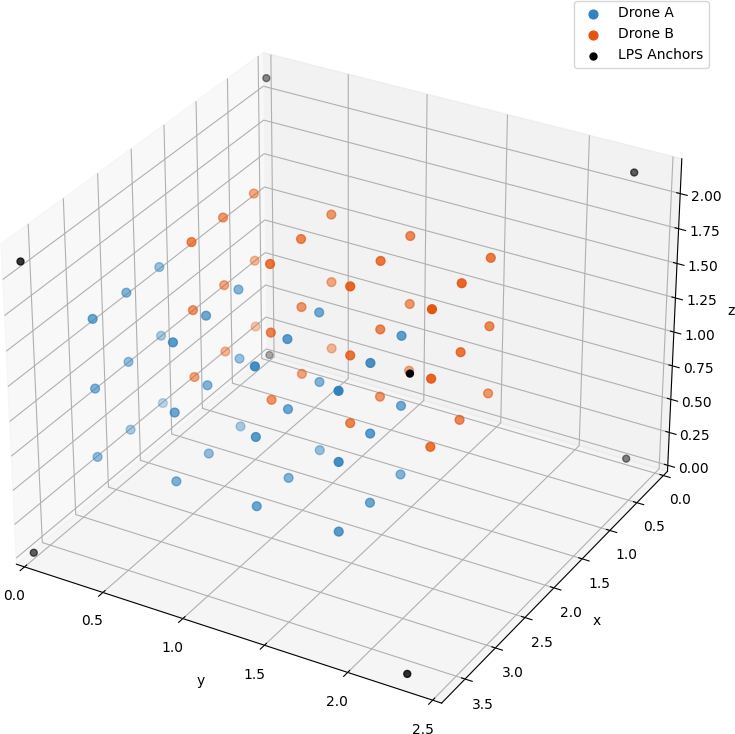}
		\caption{Scanned locations}
		\label{fig:graph_scan_locations}
	\end{center}
\end{figure}

Figure \ref{fig:graph_scan_locations} shows the distribution of the locations to scan (waypoints) for both drones. The eight black spheres represent the anchors of the Loco Positioning System.

\subsection{Client software}
The drones are controlled by a base station: a laptop running a custom Python client \cite{drone-based-rem-client} developed during project 1 which is able to communicate with the Crazyflie drones using the Python library provided by Bitcraze. 

The client is responsible for sending the drones from waypoint to waypoint and instructing them to scan. Once a scan at a waypoint is finished, the drone will send the results back to the client where they are parsed, enriched with a timestamp and stored for further processing. 

The collected samples will be $\langle timestamp, x, y, z, ssid, rssi, mac, channel \rangle$ tuples where the timestamp is set by the client upon receiving the other tuple elements from the drone.

For this project, the client was modified to be able to control multiple drones in a sequential fashion with a matching set of waypoints and parameters: radio address, starting position and yaw. While this project shows that working for two drones, it can be easily scaled up to many more by simply adding sets of waypoints and parameters. This keeps the added complexity of introducing an extra drone small and constant.

\section{Data processing}
\subsection{Exploration}
Using this setup, data was collected for further analysis and processing. A total of 2696 samples were collected, 1495 by drone A and 1201 by drone B. During data collection, drone A was active for 5 minutes 3 seconds and drone B for precisely 5 minutes.

Table \ref{tab:data_analysis_interesting_characteristics} shows a few interesting characteristics on the collected samples.

\begin{table}[H]
	\begin{center}
		\begin{tabular}{| l | c |}
			\hline
			Characteristic & Value \\ \hline
			\hline
			$\text{Samples}$ & $2696$ \\ \hline
			$\text{Distinct MAC addresses}$ & $73$ \\ \hline
			$\text{Distinct SSIDs}$ & $49$ \\ \hline
			$\text{Distinct channels}$ & $5$ \\ \hline
			$\text{Mean RSSI}$ & $-72.84$ \\ \hline 
			$\text{Median RSSI}$ & $-75$ \\ \hline 
		\end{tabular}
	\end{center}
	\caption{Characteristics of collected samples}
	\label{tab:data_analysis_interesting_characteristics}
\end{table}

As mentioned in \ref{subsec:test-environment}, the data was collected in a big apartment building, this explains why there are many more MAC addresses than SSIDs as the major ISPs are advertising their own networks on the routers/modems provided by them. These SSIDs like \textit{TelenetWiFree} are broadcasted by multiple devices.

\begin{figure}[H]
	\begin{center}
		\includegraphics[width=\columnwidth]{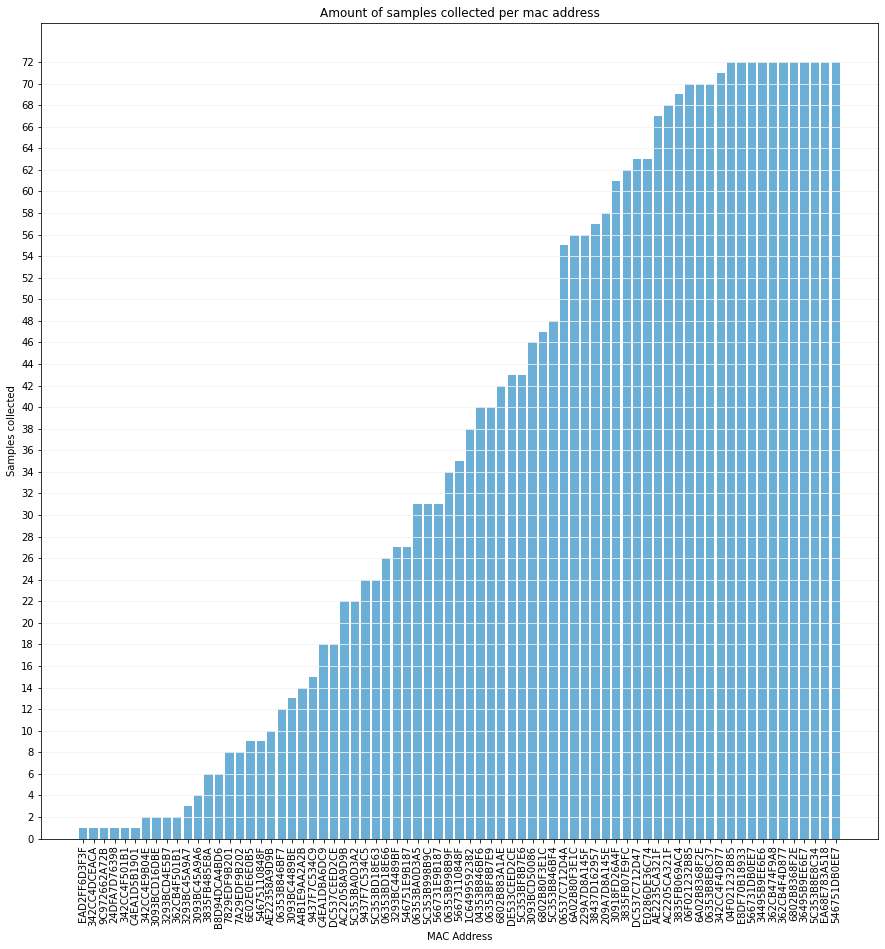}
		\caption{Amount of samples collected per MAC address}
		\label{fig:graph_samples_per_ap}
	\end{center}
\end{figure}

Figure \ref{fig:graph_samples_per_ap} shows that some access points (MAC addresses) were only seen in a few locations while 11 access points were seen in all 72 scanned locations.

\begin{figure}[H]
	\begin{center}
		\includegraphics[width=\columnwidth]{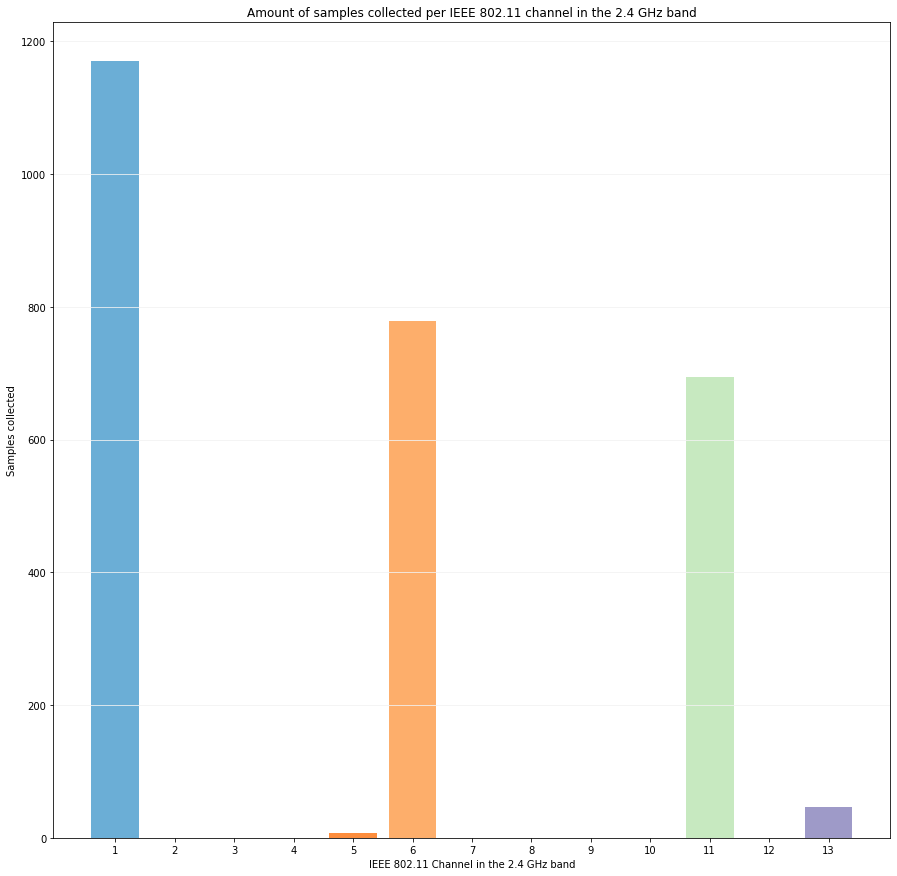}
		\caption{Amount of samples collected per IEEE 802.11 channel in the 2.4 GHz band}
		\label{fig:graph_samples_per_channel}
	\end{center}
\end{figure}

In figure \ref{fig:graph_samples_per_channel} we see that the majority of the samples were collected in just three IEEE 802.11 channels: 1, 6 and 11.

\subsection{Difference in collected samples between the drones}
As we saw in the previous section, there is a relatively big difference in samples collected between drone A and drone B, this is unexpected and warrants further investigation.

\begin{figure}[H]
	\begin{center}
		\includegraphics[width=\columnwidth]{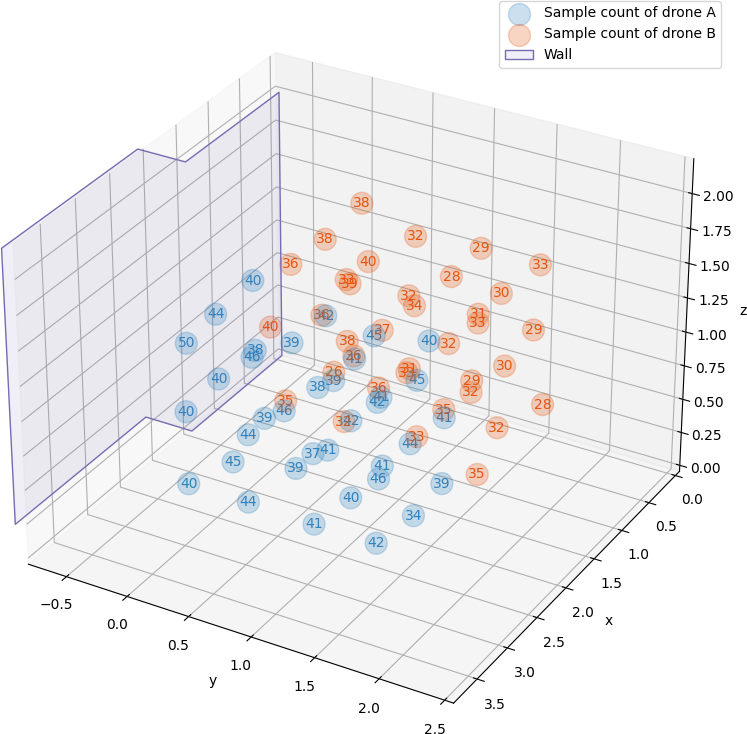}
		\caption{Amount of samples per drone and scanned location}
		\label{fig:graph_samples_per_scanned_location_per_drone.png}
	\end{center}
\end{figure}

When we look at the samples collected per drone and scanned location (figure \ref{fig:graph_samples_per_scanned_location_per_drone.png}), we see no obvious issues with the amount of samples collected by drone B, except that the amount of samples in general seems to be lower than for drone A. There are environmental factors that can play a role however:
\begin{itemize}
	\item The positive x-axis and negative y-axis point towards the centre of the apartment building where we can expect to see more signals.
	\item There is a wall segment that is 40 cm wider where drone B's measurements are taken compared to drone A, as illustrated in figure \ref{fig:graph_samples_per_scanned_location_per_drone.png}.
\end{itemize}

\begin{figure}[H]
	\begin{center}
		\includegraphics[width=\columnwidth]{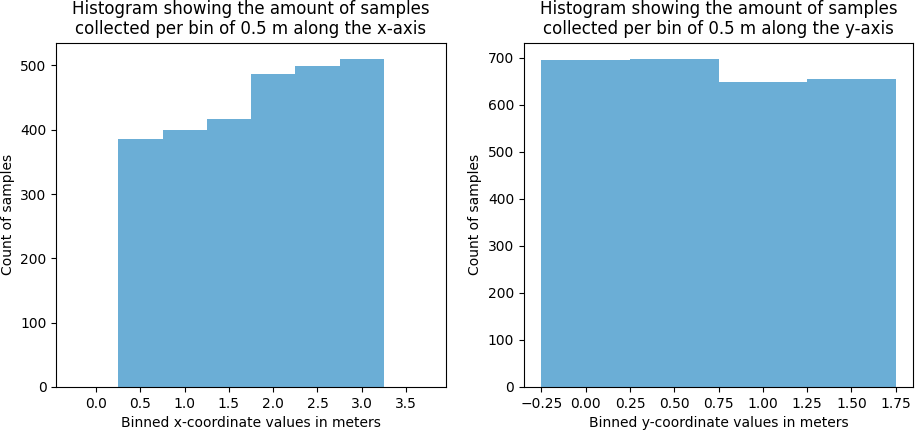}
		\caption{Histograms showing the amount of samples collected per bin of 0.5 m along the x and y-axis}
		\label{fig:graph_histogram_x_samples}
	\end{center}
\end{figure}

When we expect to see more signals (more access points) towards the centre of the building, then we should see that increase gradually, irrespective of which drone collected the sample. An illustration of this can be seen in figure \ref{fig:graph_histogram_x_samples} which shows a histogram per axis that groups the x and y values in bins of 0.5 m with the height representing the amount of samples collected by the drones in that bin. We can clearly see that the amount of samples collected increases with an increasing x-coordinate and a decreasing y-coordinate.

\subsection{Pre-processing}
A few pre-processing steps have been taken before continuing with the data:
\begin{enumerate}
	\item Since SSIDs can be shared between devices, they are not that useful and aren't used. Where appropriate, signals will be grouped based on their MAC address.
	\item The timestamps are left out of consideration as well. The time difference between the first and last collected sample is less than 10 minutes.
	\item MAC addresses with less than 16 samples will be dropped (See figure \ref{fig:graph_samples_per_ap}). While this number is arbitrary, it's not unreasonable since:
	\begin{itemize}
		\item The goal of the project is to predict RSSI values of access points for which we have measurements
		\item Enough data points per MAC address are required in order to build a reliable model. The data needs to be split up into a train / test set and potentially a validation set as well.
		\item There is a sufficient amount of MAC addresses with (close to) the maximum of 72 samples.
	\end{itemize}
	\item MAC and channel features will be considered categorical and one-hot encoded
\end{enumerate}

This pre-processing results in 2565 retained samples (131 dropped) with the features and types as illustrated in table \ref{tab:data_analysis_interesting_characteristics}:
\begin{table}[H]
	\begin{center}
		\begin{tabular}{| l | c |}
			\hline
			Feature & Type \\ \hline
			\hline
			$\text{x}$ & $\text{float64}$ \\ \hline
			$\text{y}$ & $\text{float64}$ \\ \hline
			$\text{z}$ & $\text{float64}$ \\ \hline
			$\text{rssi}$ & $\text{int64}$ \\ \hline
			$\text{mac}$ & $\text{object (one-hot encoded)}$ \\ \hline 
			$\text{channel}$ & $\text{object (one-hot encoded)}$ \\ \hline 
		\end{tabular}
	\end{center}
	\caption{Features and types of pre-processed samples}
	\label{tab:data_analysis_interesting_characteristics}
\end{table}

\subsection{Loss function}
For this regression problem, the accuracy of estimators will be measured based on the root mean square error of their predictions.

\subsection{Train / Test split}
In order to have an unbiased view on an estimator's predictive capacity, the pre-processed data will be split into a training ($75\%$) and test ($25\%$) set. For those estimators that require an additional validation set for tuning their hyperparameters, the validation set will be taken out of the training set.

\subsection{Baseline estimator}
In order to assess more elaborate estimators we'll use a baseline estimator that always returns the mean. The \verb|DummyRegressor| class of the scikit-learn package was used for this with the \verb|strategy| set to "mean" (\textit{mean} as strategy performs slightly better than \textit{median}). Running this regressor on the whole dataset as-is yields a root mean square error (RMSE) of $11.0791$ dBm. However, taking the mean RSSI over the whole dataset is not the best approach since we can expect the RSSI values of a single access point to be close to each other, but not necessarily close to (all) other access points.

The baseline was therefore adjusted to generate a regressor per MAC address, that way it will return the mean per access point. This resulted in a reduction of the error, with an RMSE now of $4.8107$ dBm. This error will be used to compare other estimators to.

\subsection{k-Nearest Neighbors estimator}
Our data is very locational as it represents signals in a 3D space, a k-nearest neighbour regressor seems therefore interesting.

The K-nearest neighbour regressor was implemented using the \verb|KNeighborsRegressor| of the sci-kit learn library. As features the x, y, z coordinates were chosen as well as the one-hot encoded MAC addresses. Including the one-hot encoded MAC addresses has the advantage that samples with a different MAC address will be considered farther away than similar samples with the same MAC address.

The kNN regressor was configured to use Euclidean distance by setting \textit{metric}=minkowski and \textit{p}=2. Euclidean distance makes sense since we have $x$, $y$ and $z$ coordinates in a three-dimensional space. The \textit{weights} and \textit{n\_neighbors} parameters were tuned using a grid search where the optimal values were \textit{weights=distance} and \textit{n\_neighbors=5}.

This resulted in an RMSE of $4.5920$ dBm, slightly better than the baseline.

As mentioned earlier, the one-hot encoded MAC addresses play an important role, a sample with a different MAC would have distance
$$ d_{s_1,s_2} = (x_{s_1}-x_{s_2})^2 + (y_{s_1}-y_{s_2})^2 + (z_{s_1}-z_{s_2})^2 + (1-0)^2 + (0-1)^2 $$
with the latter 2 terms coming from different values in the one-hot encoded columns. A sample from the same access point would have distance
$$ d_{s_1,s_3} = (x_{s_1}-x_{s_3})^2 + (y_{s_1}-y_{s_3})^2 + (z_{s_1}-z_{s_3})^2 $$
because their one-hot encoded MAC columns would completely match.

When the \verb|KneighborsRegressor|'s \textit{weights} parameter is set to "distance", RSSI values of neighbours are weighed by the distance to them. It would be interesting to have samples with a different MAC address even farther away than what the $(1-0)^2+(0-1)^2=2$ terms are currently contributing. This can be achieved by multiplying the one-hot encoded values by a chosen factor.

The optimal value for this factor was calculated by doing a grid search on values between 1 and 20. Based on the training set, performance was best when using a factor of 3 leading to an RMSE of $4.4186$ dBm. This grid search also tuned the optimal value of the \textit{n\_neighbours} parameter upwards to 16.

\subsection{k-Nearest Neighbors estimator per MAC address}
Instead of giving samples with a different MAC address a greater distance, we can also apply the same technique as was done for the baseline estimator: build a k-nearest neighbours estimator per MAC address.

We keep the hyperparameters of these MAC-based regressors the same as in the previous section but exclude the one-hot encoded MAC addresses since we're building a regressor per MAC address. While the result with an RMSE of $4.4267$ dBm is quite close the previous estimator, an improvement was expected as there is no reason taking samples of unrelated MAC addresses into account would yield a better result. This collection of regressors can only work with a small subset of the samples per regressor though, which might explain the lack in performance.

\subsection{Neural Network}
The last solution to this regression problem is to build a neural network that can predict RSSI values of our test set. The Keras library was used to build and test the network and different solutions and configurations were considered, including:
\begin{itemize}
	\item Multiple hidden layers with a varying amount of nodes
	\item Normalized RSSI values
	\item Multiple inputs: 1 for the x, y, z coordinates and 1 for the hot-encoded MAC addresses that get combined into a common hidden layer
	\item Different activation functions and optimizers
\end{itemize}

While many of these solutions had a competitive RMSE when ran against the test set, a simple neural network with a single hidden layer of 16 nodes outperformed all with an RMSE of 4.4870 dBm. While this is quite a bit better compared to our baseline, it does fall short of the best k-nearest neighbours solution discussed earlier.

This best performing neural network had the following configuration:
\begin{itemize}
	\item An input layer for the x, y, z coordinates and the one-hot encoded MAC addresses
	\item A sigmoid activation function
	\item A hidden layer with 16 fully connected nodes
	\item A linear activation function
	\item An output layer with a single node for the prediction
	\item An Adam-based optimizer
\end{itemize}

\subsection{Comparison}
\begin{figure}[H]
	\begin{center}
		\includegraphics[width=\columnwidth]{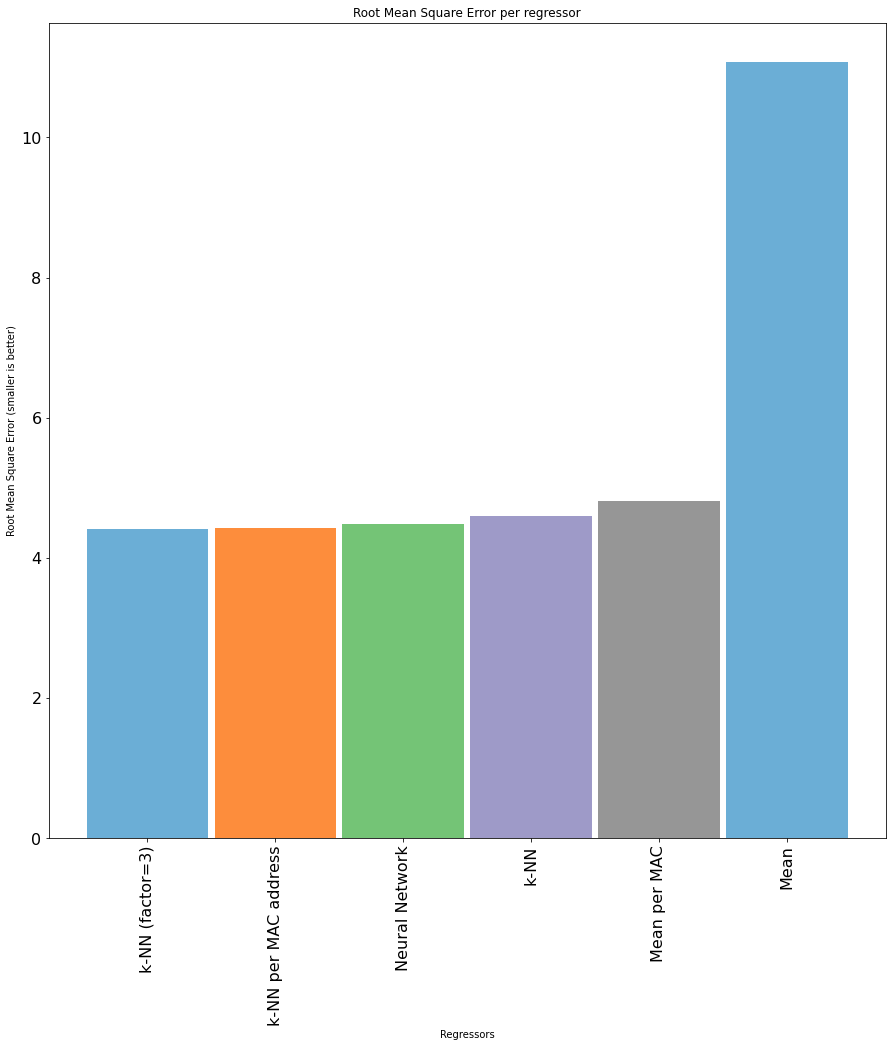}
		\caption{Histograms showing the amount of samples collected per bin of 0.5 m along the x and y-axis}
		\label{fig:root_mean_square_error_per_regressor}
	\end{center}
\end{figure}

Figure \ref{fig:root_mean_square_error_per_regressor} shows a comparison of the RMSEs of the different regressors that were tested. While the regressor that always predicted the mean RSSI of the training set didn't perform well, the results of the other regressors are close to each other, making it difficult to propose one over the other.

\section{Future Work}
While this project has shown how drones can be used to gather IEEE 802.11b, 802.11g and 802.11n beacon data at scale with the purpose of building a radio environment map, several opportunities for expansion remain.
The drones together with the base station record timestamps with the data they gather. Exploring how the data changes and how the different regressors behave with data spanning multiple days or even weeks would add an extra dimension.
Another interesting expansion would be to replace the custom ESP8266 deck with another lightweight sensor so that this system can be repurposed for collecting different types of data.
Finally, Bitcraze - the company that designed the open Crazyflie platform - has finalized  development on a new compatible infrared-based positioning system called \textit{Lighthouse}. The range of this positioning system is smaller but its accuracy and precision are competitive while requiring less anchors and being cheaper overall. This could allow for an easier to deploy solution.

\section{Conclusion}
This project has shown how the setup with an accurate positioning system, a customized drone and a controlling base-station can be used at scale to build radio environment maps. Fully autonomous, it can gather thousands of IEEE 802.11b, 802.11g and 802.11n beacon data points in the 2.4 GHz band using multiple drones and with improved stability in a real-world environment.

Several methods are proposed in this report that enable us to predict RSSI values at unknown locations with a root mean square error smaller than 4.5 dBm, outperforming the baseline mean-based predictor.

\section{Acknowledgement}
I would like to thank my supervisor for this project, Dr. Filip
Lemic, for his guidance and support over the course of this project. I would
also like to thank Prof. Dr. Jeroen Famaey for giving me the
opportunity to continue my research in this area.

Crazyflie related schematics and images (Figures 2 and 3) are copyright Bitcraze AB and are used according to their
Creative Commons Attribution 3.0 License.

\bibliographystyle{IEEEtran}

\begin{thebibliography}{00}
\bibitem{project1} Ken Mendes, Filip Lemic, and Jeroen Famaey. 2020. "Automated, autonomous, and repeatable wireless experimentation in heterogeneous 3D environments: demo abstract." 2020 In Proceedings of the 18th Conference on Embedded Networked Sensor Systems (SenSys '20). Association for Computing Machinery, New York, NY, USA, 601–602. DOI: \url{https://doi.org/10.1145/3384419.3430412}

\bibitem{crazyflie-2.1} Bitcraze Crazyflie 2.1 \\ \url{https://www.bitcraze.io/products/crazyflie-2-1/}

\bibitem{esp8266-cf-firmware} ESP-01 driver for the Crazyflie 2.1 \\
\url{https://github.com/mendesk/crazyflie-firmware-esp8266}

\bibitem{crazyflie-firmware-2021.06} Bitcraze Crazyflie Firmware release 2021.06 \\
\url{https://github.com/bitcraze/crazyflie-firmware/tree/2021.06}

\bibitem{drone-based-rem-client} Custom drone based REM client \\
\url{https://github.com/mendesk/drone-based-rem}

\bibitem{lighthouse} Bitcraze Lighthouse Positioning System \\
\url{https://www.bitcraze.io/documentation/lighthouse/}

\end{thebibliography}

\end{document}